\documentstyle[12pt]{article}
\begin{document}
\begin{center}
{\bf On the Role of Locality Condition in Bell's Theorem}\\
\vspace{0.5cm}

{\bf Habibollah Razmi}\\
{\it Department of Physics, Faculty of Sciences,
University of Qom , P.O.Box 37135-147, Qom, I.R. Iran.}\\
razmi@qom.ac.ir\\
\vspace{0.75cm}
Abstract
\end{center}

For a special stochastic realistic model in certain spin-correlation
experiments and without imposing the locality condition, an inequality
is found. Then, it is shown that quantum theory is able (is possible)
to violate this inequality. This shows that, irrelevance of
the locality condition, the quantum entanglement of the spin singlet-state
is the reason for the violation of Bell's inequality in Bell's theorem.\\

\newpage
Locality condition has an important role in many models of Bell's theorem.
In Bell's original work [1], CH model [2], and almost all the other versions,
a condition often used to derive different forms of Bell's inequality
is the so-called locality condition. In fact, in many versions
of Bell's theorem, in addition to realism (hidden variable(s)), at least, the locality condition
is considered as a necessary assumption. Although in stochastic realistic models (e.g. CH model)
the condition of determinism is not present, the so called locality condition
is still one of the necessary assumptions. Here, we want to show that even before
introducing the locality condition, quantum theory is in conflict with realistic
stochastic models.\\
Consider a source which emits a pair of spin $\frac{1}{2}$ identical particles
in a singlet-state.
The two particles of the
pair travel in opposite directions towards suitable measuring devices $D_A$ and 
$D_B$.
Identify each pair of escaping particles by an index $i (i=1,2,...,n)$.
The
experimental arrangement is such that one particle from each pair will enter
$D_A$, and the other one will enter $D_B$.
Each detector has a preferred direction. The azimuthal orientation of this
direction denoted by $Q_A$ and $Q_B$ according to whether one deals with ${\theta}_A$
or ${\theta}_B$. Denote by $r_{ji}$ the result obtained by $D_j$ in the $i$th pair.
This can take values +1 or -1 according to whether the deflection is along the
preferred direction or its opposite. Each detector can take two possible values
${\theta}_j'$ and ${\theta}_j''$. A crucial requirement is that the 
experimental set-up in sides A and B be spatially separated (i.e. having
space-like separation in the sense of the special theory of relativity).
There are four possible experimental specifications corresponding to the four
possible pairs $({\theta}_A',{\theta}_B')$ , $({\theta}_A',{\theta}_B'')$
, $({\theta}_A'',{\theta}_B')$ and $({\theta}_A'',{\theta}_B'')$.\\
By definition the correlation function $C(r_A,r_B)$ is the average of the
product of the results obtained by $D_A$ and $D_B$ :\\
\begin{equation}
C(r_A,r_B) = \frac{1}{n}\sum_{i=1}^n r_{Ai}({\theta}_A,{\theta}_B)r_{Bi}({\theta}_A,{\theta}_B)
\end{equation}
Usually, the following locality condition is imposed [3-4]:\\
\begin{equation}
\left\{\begin{array}{lll}
r_A({\theta}_A,{\theta}_B) = r_A({\theta}_A)\\
r_B({\theta}_A,{\theta}_B) = r_B({\theta}_B)
\end{array}\right.
\end{equation}
and the correlation function (1) is written as:\\
\begin{equation}
C(r_A,r_B) = \frac{1}{n}\sum_{i=1}^n r_{Ai}({\theta}_A)r_{Bi}({\theta}_B)
\end{equation}
and then an inequality (e.g. Bell's inequality) is found which cannot be always
satisfied by quantum mechanics (Bell's theorem).\\
Here we are going to show that even before the introduction of the locality condition
(2) (i.e. by using only the correlation function (1)), there is a descrepancy
between the model used and quantum theory (Bell's theorem without locality condition).\\
Since both $r_{Ai}$ and $r_{Bi}$ are dichotomic variables (i.e. take values +1 or -1 only),
their product
is also dichotomic. Assume that from $n$ runs of the experiment, $m$ of them
show the value +1 for the product $r_Ar_B$; thus, $n-m$ of them would show the value -1.
Therefore, the correlation function (3) is reduced to the following form:\\
\begin{equation}
C(r_A,r_B) = \frac{1}{n}([\sum_{i=+1}^m (+1)] + [\sum_{i=m+1}^n (-1)]) 
\end{equation}
Or:
\begin{equation}
C(r_A,r_B) = \frac{2m}{n} - 1 
\end{equation}
Here, we should mention that although $m$ has a stochastic and unknown value,
it is an integer between 0 and $n$ depending on the experimental arrangement
(orientation) of the devices.\\
For Future use, specially for the simplicity of
algebra, let us define the following $S$-function as:
\begin{equation}
S(r_A, r_B) = \frac{[ 1 + C(r_A, r_B)]}{2}
\end{equation}
which by means of (5), takes the form:
\begin{equation}
S(r_A, r_B) = \frac{m}{n}
\end{equation}
One may consider $N$ experminets all with $n$ runs and similar in set-up but with arbitrary
direction of detectors. Then, for each of these experiments the $S$-function is:
\begin{equation}
S_l = \frac{m_l}{n} \ \ \ \ \ \ \ \ \ \ \ \ \ \ \ l= 1, 2, ..., N
\end{equation}
If in all these experiments the special "deterministic" case which
corresponds to ${\theta}_A^l={\theta}_B^l$ is excluded then all $m_l$s satisfy:
\begin{equation}
0 < m_l \leq n
\end{equation}
This is because the special case $m_l=0$ only corresponds to setting
${\theta}_A^l={\theta}_B^l$.\\
We are interested to the following $\Gamma$-function corresponding to the results
of the $N$ experiments:
\begin{equation}
\Gamma = \sum_{l=1}^N S_l = \sum_{l=1}^N \frac{m_l}{n}
\end{equation}
A comparison of (9) and (10) results in:
\begin{equation}
\Gamma \geq \frac{N}{n}
\end{equation}
This inequality may be considered as a special kind of "Bell's inequality"
for the stochastic realistic model considered here ( with $N$ experiments each
with $n$ runs and with arbitrary direction of detection excluding the special
"deterministic" case with ${\theta}_A^l={\theta}_B^l$).\\
It is important to emphasize once again that the inequality (11) has been derived
without imposing the locality condition (2).\\
Now, the question is: Does quantum theory satisfy the inequality (11)?\\
The quantum mechanical result for the correlation function of the above-mentioned
experimental set-up (singlet-state) is [5]:
\begin{equation}
C_{QM}(r_A,r_B) = -\cos({\theta}_A-{\theta}_B)
\end{equation}
Thus:
\begin{equation}
S_{QM} (r_A, r_B) = {\sin}^2(\frac{{\theta}_A-{\theta}_B}{2}) = {\sin}^2(\frac{{\theta}_{AB}}{2})
\end{equation}
where ${\theta}_{AB} = {\theta}_A - {\theta}_B$.\\
Therefore:
\begin{equation}
{\Gamma}_{QM} = \sum_{l=1}^N {\sin}^2(\frac{{\theta}_{AB}^l}{2})
\end{equation}
This ${\Gamma}_{QM}$ is able (is possible) to violate the inequality (11).
Let show this in more detail.\\
Suppose all ${\theta}_{AB}^l$s are selected so that:
\begin{equation}
0 < {\theta}_{AB}^l< 2Arc\sin (\frac{1}{\sqrt n}) \ \ \ \ \ \ \ \ \ \ \ \ l=1,2,..,N
\end{equation}
Then:
\begin{equation}
{\Gamma}_{QM} < \frac{N}{n}
\end{equation}
This is in conflict with the inequality (11). It means that one can always
find cases for them quantum theory cannot reproduce the predictions of the
special stochastic realistic (not necessarily local) model. This can be
considered as a model of Bell's theorem without locality condition.\\
In fact, in the correlation function introduced in (1), the factorization of the terms under sum in the
form of $r_A.r_B$ ,irrelevance of the functional form of $r_A$ and $r_B$ on the parameters
of the experiment (i.e. irrelevance of the locality condition), is not possible. This
is because the singlet-state is a quantum mechanical nonfactorizable state. This mathematical
nonfactorizability of the spin singlet-state is due to its physical entanglement
which has a very important scheme in modern theory of quantum information
and computation [8].\\
Two mathematical points should be explained more about the relations (15) and
(16). The first is that if $n\rightarrow\infty$ then (15) cannot be correct.
In fact, in real experiments although $n$ may have large values, never goes to
infinity. Here, we appeal to the data of one of real experiments in proton-proton
scattering experiment of Lamehi-Rachti and Mittig [6],
which exemplifies the above set-up where the total count $(n)$ was $10^4$ [7]. The second point is that if
${\theta}_{AB}^l$s were greater than $\frac{\pi}{2}$, one could not reach
(16) from (15). Simply, we can always choose number of runnings $n$ so that
the right hand side in (15) be less than $\frac{\pi}{2}$. It is enough only to
show that quantum theory is able (is possible) to violate (11).\\
\newpage
\begin{itemize}
\item[1] J.S. Bell, Physics {\bf 1}, No.3, 195 (1964).
\item[2] J.F. Clauser and M.A. Horne, Phys. Rev. D {\bf 10}, 526 (1974).
\item[3] H.P. Stapp, Am.J. Phys. {\bf 53}, 306 (1985).
\item[4] H.P. Stapp, Found. Phys. {\bf 18}, 427 (1988).
\item[5] L.E. Ballentine, {\bf Quantum Mechanics}, 440 (Prentice-Hall, 1990).
\item[6] M. Lamehi-Rachti and W. Mittig, Phys. Rev. {\bf 14}, 2543 (1976).
\item[7] Private Communication with Dr. Lamehi-Rachti.
\item[8] D. Bouwmeester, et al, {\bf The Physics of Quantum Information},
(Springer, 2000).

\end{itemize}

\end{document}